\documentclass[aps,prd,superscriptaddress,twocolumn,floatfix,longbibliography]{revtex4-2}  

\usepackage{graphicx,color}
\usepackage[colorlinks=true,linkcolor=blue,citecolor=blue,linkcolor=blue]{hyperref}%

\usepackage{amsmath,amssymb,amsfonts,bm}     
\usepackage{mathrsfs}
\usepackage{braket}
\allowdisplaybreaks[4]

\begin{document}
	

	\title{Finite temperature Casimir effect in one-dimensional scalar field with double delta-function potentials}

	\author{Liang Chen}
	\email[Corresponding Email:]{slchern@ncepu.edu.cn}
	\affiliation{School of Mathematics and Physics, North China Electric Power University, Beijing 102206, China}
	\affiliation{Institute of Condensed Matter Physics, North China Electric Power University, Beijing 102206, China}
	\affiliation{Hebei Key Laboratory of Physics and Energy Technology, North China Electric Power University, Baoding 071003, China}	
	
	\author{Xu-Feng Zhao}
	\affiliation{School of Mathematics and Physics, North China Electric Power University, Beijing 102206, China}
	\affiliation{Institute of Condensed Matter Physics, North China Electric Power University, Beijing 102206, China}

	\author{Shao-Zhe Lu}
	\affiliation{School of Mathematics and Physics, North China Electric Power University, Beijing 102206, China}
	\affiliation{Institute of Condensed Matter Physics, North China Electric Power University, Beijing 102206, China}

\begin{abstract}
We investigate the finite-temperature Casimir effect for a (1+1)-dimensional scalar field interacting with a pair of delta-function potentials. We employ the canonical quantization method to compute the Casimir force and entropy, contrasting the results with those from the standard Lifshitz theory. At zero temperature, both frameworks yield identical forces. For the finite-temperature case, we find that in the long-distance limit, the Casimir force decays asymptotically as $F_C(a,T)=-T/(4a)$, with the Lifshitz theory predicting a magnitude twice as large as that from canonical quantization. Crucially, the canonical quantization method yields a physically consistent entropy that remains positive and increases with temperature. These results demonstrate the robustness of the canonical quantization approach in providing a thermodynamically sound description of the thermal Casimir effect in this system.
\end{abstract}

\maketitle

\section{Introduction}
The Casimir effect, a fundamental phenomenon in quantum field theory, arises from the zero-point fluctuations of quantized fields in the presence of material boundaries or external potentials. Since its original prediction by H. B. G. Casimir for perfect metal plates \cite{Casimir1948PKNAW}, this effect has been extensively studied across a variety of physical contexts, with implications spanning condensed matter physics, cosmology, and nanoscale engineering  \cite{Milonni1994QVacuumbook,Mostepanenko1997Casimirbook,Bordag2009Casimirbook,Bordag2001PhysRep}. Accurate modeling of Casimir forces has become essential for the design, operation, and control of micro- and nano-electromechanical systems, where quantum fluctuations can dominate interfacial behavior \cite{BuksE2001PRB,SerryFM1998JAP,RodriguezAW2011APL}. 

At finite temperatures, the interplay between quantum and thermal fluctuations leads to the thermal Casimir effect \cite{SushkovAO2011NatPhys}, which exhibits intriguing thermodynamic features---most notably, the emergence of negative entropy in certain parameter regimes. Early discussions of negative entropy largely attributed it to dissipative response in real metals, particularly within the Drude model framework  \cite{Mostepanenko2006JPAMG,Mostepanenko2008JPA,Klimchitskaya2008JPAMT,Lamoreaux2012ARNPS,HartmannM2017PRL,LiuM2019PRB,Klimchitskaya2022IJMPA,Mostepanenko2021Universe}. As emphasized by V. M. Mostepanenko \cite{Mostepanenko2021Universe}, this behavior is closely tied to the “Casimir puzzle”, a persistent discrepancy between experimental results and theoretical predictions based on the Drude description. However, growing evidence indicates that negative entropy is not solely a consequence of dissipation. Even in idealized systems with no dissipation---such as those described by the plasma model or perfect conductor boundary conditions---negative entropy can occur due to purely geometric effects.
A substantial body of recent work has systematically identified geometry-induced negative entropy across a range of Casimir configurations. Examples include sphere–sphere and sphere–plate geometries  \cite{RodriguezLopezP2011PRB,IngoldGL2015PRE,MiltonKA2016FdP}, spherical plasma shells  \cite{BordagM2018JPA}, free-standing thin sheets \cite{LiY2016PRD,BordagM2018PRD}, and periodic backgrounds \cite{BordagM2020EPJC}.  
These studies collectively establish that negative entropy is an intrinsic feature of the Casimir interaction in specific geometries, independent of dissipative material response. Its origin lies in the nontrivial mode structure imposed by boundaries and the distinct roles of transverse electric and transverse magnetic field components, with the transverse electric contribution often serving as the primary source of negative entropy.  These findings challenge a purely material-centered interpretation and call for a deeper understanding of thermodynamic consistency in Casimir systems.

In this work, we reexamine the finite-temperature Casimir interaction between two delta-function potential barriers using both the canonical quantization method and the Lifshitz formula. While the zero-temperature Casimir interaction in this model has been widely investigated in earlier studies  \cite{BordagM1992JPA,GrahamN2002NPB,GrahamN2003PLB,GrahamN2004NPB,MiltonKA2004JPA}, its thermodynamic behavior at finite temperature---particularly the entropy---remains less explored. 
We compute the Casimir force and entropy within the canonical quantization framework and systematically compare the results with those derived from the standard Lifshitz formula. 
The paper is organized as follows. In Sec. \ref{sec2}, we introduce the model and outline the canonical quantization procedure. Section \ref{sec3} presents the zero-temperature results, demonstrating agreement between the two methods. In Sec. \ref{sec4}, we analyze the finite-temperature behavior and discuss the entropy.  We summarize our results in Sec. \ref{sec5}. 

\section{Theoretical Model} \label{sec2}
We consider a scalar field $\phi(x,t)$ in (1+1)-dimensional spacetime governed by the equation, 
\begin{equation}
\left[\frac{\partial^2}{\partial t^2} - v^2 \frac{\partial^2}{\partial x^2} + U(x)\right] \phi(x,t) = 0,  \label{eq1}
\end{equation}
where $v$ denotes the propagation velocity of the scalar field, and the potential $U(x)$ is given by
\begin{equation}
U(x) = \gamma \left[ \delta\left(x + \frac{a}{2}\right) + \delta\left(x - \frac{a}{2}\right) \right].  \label{eq}
\end{equation}
Here, $a$ is the separation between the two delta-function potential barriers, and $\gamma$ represents their strength. The vacuum fluctuation-induced Casimir interaction between the barriers can be evaluated using the standard Lifshitz formula. At finite temperature $T$, the Helmholtz free energy reads
\begin{equation}
\mathcal{F}_L(a,T) = T \sum_{n=0}^{\infty}{}^{\prime} \log\left[ 1 - e^{-2a\zeta_n/v} \left( \frac{\gamma}{\gamma + 2v\zeta_n} \right)^2 \right], \label{eq3}
\end{equation}
where $\zeta_n=2\pi{n}{T}/\hbar$ is the Matsubara frequency at temperature $T$ and $\hbar$ is the reduced Planck's constant (we set the Boltzmann constant $k_B = 1$), the prime in the summation means that for the special case $n=0$, a prefactor $1/2$ is multipiled. The Casimir force $F(a,T)$ and the Casimir entropy $S(a,T)$ are then derived from the free energy as
\begin{gather}
F_L(a,T)=-\frac{\partial\mathcal{F}(a,T)}{\partial{a}},  \label{eq4} \\
S_L(a,T)=-\frac{\partial\mathcal{F}(a,T)}{\partial{T}}.  \label{eq5}
\end{gather}
In all these expressions, the index $L$ is used to clarify the results obtained from Lifshitz theory, distinguishing from the canonical quantization method introduced below. In the context of our current problem, two aspects of the Lifshitz theory warrant careful scrutiny. First, within the imaginary-frequency formalism, electromagnetic waves in dielectric media are treated as attenuated waves—a physically justified approach for infinite dielectric plates, where wave attenuation indeed occurs. However, for an infinitely thin delta-function barrier, applying the same attenuated-wave description to the scalar field in the external vacuum lacks a clear physical interpretation. Second, the Helmholtz free energy expression itself presents a difficulty: the zero-Matsubara-frequency term diverges, as seen from
\begin{equation}
\log\left[1 - e^{-2a\zeta_0/v} \left( \frac{\gamma}{\gamma + 2v\zeta_0} \right)^2 \right] = \infty,  \label{eq6}
\end{equation}
which renders the free energy ill-defined. This divergence necessitates careful treatment in the analysis.

In this work, we adopt a more physically transparent approach---canonical quantization---to compute the Casimir force and entropy \cite{KupiszewskaD1990PRA,KupiszewskaD1992PRA,vanEnkSJ1995PRA}. Within this framework, the Casimir force and entropy are derived from the quantized energy-momentum tensor. For a scalar field propagating from left to right (i.e., $k>0$), the wave function in different spatial regions can be written as:
\begin{gather}
\varphi_{\text{I}}(k,x<-a/2) = \frac{e^{ikx}}{\sqrt{2\pi}} + B_k \frac{e^{-ikx}}{\sqrt{2\pi}},  \label{eq7} \\
\varphi_{\text{II}}(k,-a/2<x<a/2) = C_k \frac{e^{ikx}}{\sqrt{2\pi}} + D_k \frac{e^{-ikx}}{\sqrt{2\pi}}, \label{eq8} \\
\varphi_{\text{III}}(k,x>a/2) = G_k \frac{e^{ikx}}{\sqrt{2\pi}}.  \label{eq9}
\end{gather}
By imposing continuity conditions on the wave function, the coefficients $B_k$, $C_k$, $D_k$, $G_k$ can be determined. Explicitly, we obtain:
\begin{gather}
C_k = \frac{2v^2k(2v^2k + i\gamma)}{(2v^2k + i\gamma)^2 + e^{2ika}\gamma^2}, \label{eq10} \\
D_k = -\frac{2ie^{ika}\gamma v^2 k}{(2v^2k + i\gamma)^2 + e^{2ika}\gamma^2}.  \label{eq11}
\end{gather}
Moreover, direct calculation confirms that the probability flow remains continuous, as reflected by the identities: 
\begin{gather}
|B_k|^2+|G_k|^2=1, \label{eq12} \\
|D_k|^2+|G_k|^2=|C_k|^2.  \label{eq13}
\end{gather}
For left-propagating modes ($-k<0$), the corresponding expressions are:
\begin{gather}
\varphi_{\text{I}}(k,x<-a/2) = G_{-k} \frac{e^{-ikx}}{\sqrt{2\pi}}, \label{eq14} \\
\varphi_{\text{II}}(k,-a/2<x<a/2) = C_{-k} \frac{e^{-ikx}}{\sqrt{2\pi}} + D_{-k} \frac{e^{ikx}}{\sqrt{2\pi}}, \label{eq15}  \\
\varphi_{\text{III}}(k,x>a/2) = \frac{e^{-ikx}}{\sqrt{2\pi}} + B_{-k} \frac{e^{ikx}}{\sqrt{2\pi}}. \label{eq16}
\end{gather} 
A detailed calculation shows that $C_{-k}=C_k$ and $D_{-k}=D_k$. The field operator $\hat{\phi}(x,t)$ and its conjugate momentum $\hat{\pi}(x,t)=\dot{\hat{\phi}}(x,t)$ are quantized as: 
\begin{widetext}
\begin{gather}
\hat{\phi}(x,t) = \int_0^{\infty} \frac{\mathrm{d}k}{\sqrt{2\pi}} \sqrt{\frac{\hbar}{2\omega}} \left[ \hat{a}(k)\varphi(k,x) + \hat{b}(-k)\varphi(-k,x) \right] e^{-i\omega t} + {h.c.},  \label{eq17} \\
\hat{\pi}(x,t) = -i \int_0^{\infty} \frac{\mathrm{d}k}{\sqrt{2\pi}} \sqrt{\frac{\hbar\omega}{2}} \left[ \hat{a}(k)\varphi(k,x) + \hat{b}(-k)\varphi(-k,x) \right] e^{-i\omega t} + {h.c.},  \label{eq18}
\end{gather}
\end{widetext}
where $\omega=ck$ is the eigenfrequency of the mode, and $h.c.$ denotes the Hermitian conjugate. The annihilation operators $\hat{a}(k)$, $\hat{b}(-k)$ and their conjugates, creation operators, $\hat{a}^{\dag}(k)$, $\hat{b}^{\dag}(-k)$ satisfy the standard commutation relations:
\begin{gather}
[\hat{a}(k),\hat{a}^{\dag}(k^{\prime})]=\delta(k-k^{\prime}),  \label{eq19} \\
[\hat{b}(-k),\hat{b}^{\dag}(-k^{\prime})]=\delta(k-k^{\prime}).  \label{eq20}
\end{gather}

In the framework of classical physics, the energy-momentum tensor of the scalar field is given by
\begin{equation}
\mathcal{T}^{\mu\nu} = (\partial^\mu \phi)(\partial^\nu \phi) - \frac{1}{2} \eta^{\mu\nu} (\partial_\rho \phi \partial^\rho \phi), \label{eq21}
\end{equation}
and satisfies the energy-momentum conservation law
\begin{equation}
\partial_\mu \mathcal{T}^{\mu\nu} = 0. \label{eq22}
\end{equation}
For the (1+1)-dimensional scalar field considered here, the metric tensor is $\eta^{\mu\nu}=\text{diag}(1,-v^2)$, and the energy-momentum tensor takes the explicit form, 
\begin{equation}
\mathcal{T}=\begin{bmatrix}
\frac{1}{2}\left[(\partial_t \phi)^2 + v^2 (\partial_x \phi)^2 \right] & -v^2 (\partial_t \phi)(\partial_x \phi) \\
-v^2 (\partial_t \phi)(\partial_x \phi) & \frac{v^2}{2} \left[ (\partial_t \phi)^2 + v^2 (\partial_x \phi)^2 \right]
\end{bmatrix}. \label{eq23}
\end{equation}
The conservation equation of the linear momentum density, $-(\partial_{t}\phi)(\partial_{x}\phi)$, reads
\begin{equation}
\partial{t}\mathcal{T}^{0,x}=-\partial{x}\mathcal{T}^{x,x}.  \label{eq24}
\end{equation}
This implies that the force exerted on the right $\delta$-function barrier at $x=a/2$ is given by
\begin{equation}
F_C(a) = -\frac{1}{v^2} \left( \mathcal{T}^{xx} \big|_{x = \frac{a}{2}+0^+} - \mathcal{T}^{xx} \big|_{x = \frac{a}{2}+0^-} \right). \label{eq25}
\end{equation}
Here the subindex $C$ is used to clarify the Casimir force obtained from canonical quantization. Upon quantizing the field, the Casimir force at zero temperature becomes
\begin{equation}
F_C(a) = -\bra{0} \left[ (\partial_t \hat{\phi})^2 + v^2 (\partial_x \hat{\phi})^2 \right] \ket{0}, \label{eq26}
\end{equation}
where the field operator $\hat{\phi}$ is defined  in Eq. (\ref{eq17}). Using the explicit expressions for the eigenmodes in Eqs. (\ref{eq7})-(\ref{eq9}) and (\ref{eq14})-(\ref{eq16}), together with the probability-flow continuity conditions in Eqs. (\ref{eq12}) and (\ref{eq13}), we obtain 
\begin{equation}
F_C(a) = \int_0^\infty \frac{\mathrm{d}k}{2\pi} \hbar v k \left( |C_k|^2 + |D_k|^2 - 1 \right). \label{eq27}
\end{equation}
At finite temperature $T$, the Casimir force is given by
\begin{equation}
F_C(a,T) = -\int_0^\infty \mathrm{d}k P_n(k) \bra{n_k} \left[ (\partial_t \hat{\phi})^2 + v^2 (\partial_x \hat{\phi})^2 \right] \ket{n_k},  \label{eq28}
\end{equation}
where, $P_n(k)=\sum_{n=0}^{\infty}\frac{e^{-n\hbar{vk}/T}}{Z(k)}$ is the probability of finding $n$ particles in the mode with wave number $k$, and $Z(k)=\sum_n{e^{-n\hbar{vk}/T}}$ is the corresponding partition function. A straightforward calculation then yields
\begin{equation}
F_C(a,T) = \int_0^\infty \frac{\mathrm{d}k}{2\pi} \frac{\hbar v k}{1-e^{-\hbar v k / T}} \left( |C_k|^2 + |D_k|^2 - 1 \right).  \label{eq29}
\end{equation}

\section{Casimir Force at zero temperature} \label{sec3}
\begin{figure}[tb]
	\centering
	\includegraphics[width=\linewidth]{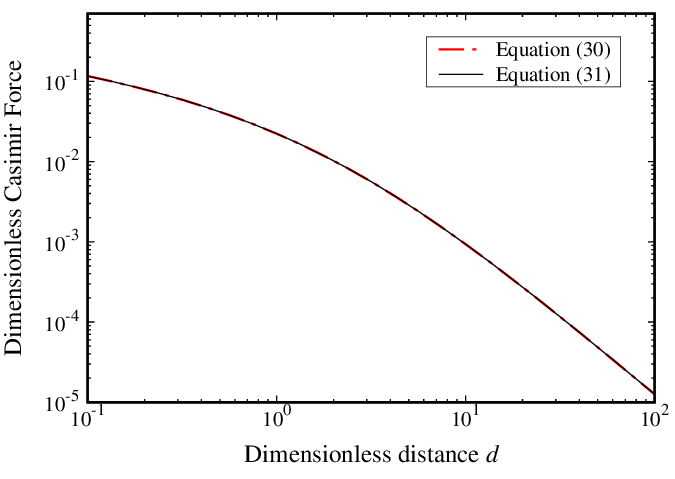}
	\caption{Casimir force in units of $-\hbar\gamma^2/v^3$ vs dimensionless distance $d=\gamma{a}/v^2$. }
	\label{fig_1}
\end{figure}

Using the explicit forms of $C_k$ and $D_k$ given in Eqs. (\ref{eq10}) and (\ref{eq11}), the zero-temperature Casimir force can be expressed as 
\begin{equation}
F_C(a) = -\frac{\hbar \gamma^2}{v^3} \int_0^\infty \frac{\mathrm{d}q}{2\pi} q \left[ 1 - \frac{8q^2 (1 + 2q^2)}{|1 - e^{2 i d q} (1 + 2 i q)^2|^2} \right], \label{eq30}
\end{equation}
where we have introduced the dimensionless wave number $q=v^2k/\gamma$ and dimensionless separation $d=\gamma{a}/v^2$. A natural question is whether this result, derived via canonical quantization, agrees with that obtained from the standard Lifshitz formula. The zero-temperature Casimir force from the latter reads
\begin{equation}
F_L(a) = -\frac{\hbar \gamma^2}{v^3} \int_0^\infty \frac{\mathrm{d}\zeta}{4\pi} \frac{\zeta}{e^{d \zeta} (1 + \zeta)^2 - 1}.  \label{eq31}
\end{equation}
We note that directly relating the two expressions through analytic continuation is nontrivial.  The denominator in Eq. (\ref{eq30}) shows that, upon continuing $q$ into the complex plane, the integrand exhibits infinitely many singularities in both the upper and lower half-planes, which obstructs a straightforward Wick rotation to the imaginary axis. Nevertheless, the integral in Eq. (30) is convergent---as $q\rightarrow\infty$, the integrand behaves asymptotically as $\cos(2dq)/(2q)$, leading to a well-defined result expressible in terms of cosine integrals. High-precision numerical integration was performed to compare Eqs. (\ref{eq30}) and (\ref{eq31}). As shown in Fig. \ref{fig_1}, the results from both methods coincide within numerical precision. Although a rigorous analytic proof of the equivalence remains open, this numerical agreement strongly supports the validity of the canonical quantization approach adopted in this work. 

\section{Casimir Force and Casimir Entropy at finite temperature} \label{sec4}
The consistency of our zero-temperature results supports the validity of the canonical quantization approach. We now extend the analysis to finite temperature. Using Eqs. (\ref{eq10}) and (\ref{eq11}), the Casimir force at temperature $T$ can be written in dimensionless form as
\begin{gather}
F_C(a,T)=-\frac{\hbar\gamma^2}{{v^3}}\int_0^{\infty}\frac{dq}{2\pi} \frac{q}{1-e^{-q/\mathsf{T}}}\notag \\
\times\left[1-\frac{8q^2(1+2q^2)}{|1-e^{2idq}(1+2iq)^2|^2}\right],  \label{eq32}
\end{gather}
where $\mathsf{T}=vT/(\hbar\gamma)$ is the dimensionless temperature, $q=v^2k/\gamma$ is the dimensionless wave number, and $d=\gamma{a}/v^2$ is the dimensionless plate separation. In the Lifshitz approach, special care is required for the zero-frequency ($n=0$) term in Eq. (\ref{eq3}). Introducing an infrared cutoff $2\pi{T}/{\hbar\Lambda}$, where $\Lambda\rightarrow\infty$ is a dimensionless parameter, the free energy can be regularized as
\begin{gather}
\mathcal{F}_L(a,T)=T\sum_{n=1}^{\infty}\log\left[1-e^{-2a\zeta_n/v}\left(\frac{\gamma}{\gamma+2v\zeta_n}\right)^2\right] \notag \\
+\frac{T}{2}\log\left[\frac{2\pi{T}}{\hbar\Lambda}\left(\frac{a}{v}+\frac{2v}{\gamma}\right)\right], \label{eq33}
\end{gather}
Using Eq. (\ref{eq4}) and converting to dimensionless variables, the Casimir force from the Lifshitz theory becomes
\begin{equation}
F_L(a,T)=-\frac{\hbar\gamma^2}{{v^3}}\left[\sum_{n=1}^{\infty}\frac{4\pi{n}\mathsf{T}^2}{e^{4\pi{n}\mathsf{T}{d}}(1+4\pi{n}\mathsf{T})^2-1}+\frac{\mathsf{T}}{2(d+2)}\right]. \label{eq34}
\end{equation}

\begin{figure}[tb]
	\centering
	\includegraphics[width=\linewidth]{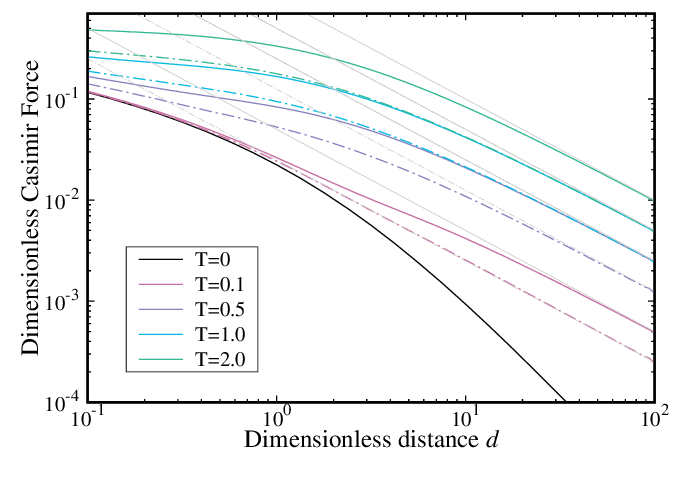}
	\caption{Casimir force in units of $-\hbar\gamma^2/(4\pi{v^3})$ vs dimensionless distance $d=\gamma{a}/v^2$. The solid lines show results from the Lifshitz theory, Eq. (\ref{eq34}). The dotted lines show results from canonical quantization method, Eq. (\ref{eq32}). }
	\label{fig_2}
\end{figure}

Notably, the final expression for $F_L(a,T)$ is independent of the infrared cutoff $\Lambda$, allowing a direct comparison with the canonical quantization result. Figure \ref{fig_2} shows the finite-temperature Casimir force as a function of dimensionless distance $d$. It is clear that, unlike the zero-temperature case, the two methods now yield different results: the attractive force from canonical quantization is smaller than that from the Lifshitz theory. In the long-distance regime ($\mathsf{T}\gg1$), both forces decay as $1/d$. Specifically, as $d\rightarrow\infty$, the Lifshitz result asymptotically behaves as
\begin{equation}
F_L(a,T)=-(\hbar\gamma^2/2v^3)\mathsf{T}/d=-T/2d, \label{eq35}
\end{equation}
whereas the canonical quantization gives
\begin{equation}
F_C(a,T)=-(\hbar\gamma^2/4v^3)\mathsf{T}/d=-T/4d. \label{eq36}
\end{equation}
This $1/d$ scaling has an important implication: if the free energy is defined to be zero at infinite separation, then integrating the Casimir force leads to a logarithmic divergence in the infrared limit. Both methods exhibit this behavior, suggesting that the logarithmic divergence is an intrinsic property of the system. Although the Casimir force itself remains finite, the divergence in the free energy will necessarily appear in other thermodynamic quantities, such as the Casimir entropy. 

The Casimir entropy can be derived from the Maxwell's relation \cite{CallenHB1985book}, i.e., 
\begin{equation}
S_C(a,T)=S_C(\infty,T)-\int_{a}^{\infty}\frac{\partial{F}_C(\tilde{a},T)}{\partial{T}}\mathrm{d}\tilde{a}, \label{eq37}
\end{equation}
If the Casimir entropy is defined to vanish at infinite separation, $S_C(\infty,T)=0$, we obtain 
\begin{equation}
S_C(a,T)=-\int_{a}^{\infty}\frac{\partial{F}_C(\tilde{a},T)}{\partial{T}}\mathrm{d}\tilde{a},  \label{eq38}
\end{equation}
which provides a practical approach for calculating and measuring the Casimir entropy. Substituting  Eq. (\ref{eq32}) into this expression yields
\begin{gather}
S_C(a,T)=\int_d^{\infty}\mathrm{d}\tilde{d}\int _0^{\infty}\frac{\mathrm{d}q}{2\pi}\left(\frac{q}{2\mathsf{T}}\right)^2\text{csch}^2\left(\frac{q}{2\mathsf{T}}\right) \notag \\
\times
\left[1-\frac{8q^2(1+2q^2)}{|1-e^{2i\tilde{d}q}(1+2iq)^2|^2}\right].  \label{eq39}
\end{gather}

\begin{figure}[tb]
	\centering
	\includegraphics[width=\linewidth]{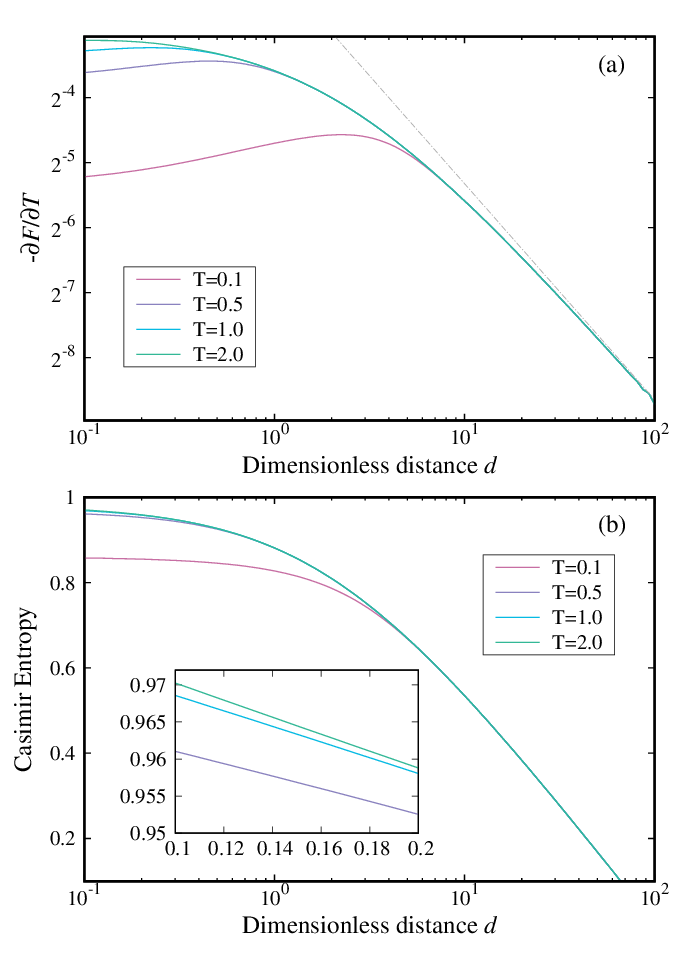}
	\caption{Panels (a) and (b) display the integrand and the resulting Casimir entropy, respectively, versus the separation distance for different temperatures. The inset shows a detailed view of the short-separation regime. }
	\label{fig_3}
\end{figure}

It is important to note that the integration over distance is infrared-divergent, which ensures the positivity of the entropy. In numerical evaluations, an infrared cutoff must be introduced. Figures \ref{fig_3}(a) and \ref{fig_3}(b) show the integrand $-\partial{F}_C/\partial{T}$ and Casimir entropy, respectively. In the long-distance limit, the integrands for different temperatures approach the universal curve $1/(4d)$. In our numerical integration, we set the infrared cutoff to $\Lambda=10^2$. Compared to the Lifshitz theory, the canonical quantization method offers clearer physical insights into the thermal Casimir effect. In the high-temperature and long-distance regime, Lifshitz theory predicts that the free energy is dominated by the zero-frequency term $(T/2)\log[2\pi{T}(a/v+2v/\gamma)/(\hbar\Lambda)]$, leading to the asymptotic expression for the Casimir entropy:
\begin{equation}
S_L(a\gg{v^2/\gamma},T)\approx-\frac{1}{2}\log\left[\frac{2\pi\mathsf{T}}{\Lambda}(d+2)\right]-\frac{1}{2},   \label{eq40}
\end{equation}
Similar to Eq. (\ref{eq39}), this expression is positive and infrared-divergent as $\Lambda\rightarrow+\infty$. However, when examining the temperature dependence of the entropy, the Lifshitz theory predicts a decrease in entropy with increasing temperature: 
\begin{equation}
\frac{\partial{}S_L(a\gg{v^2/\gamma},T)}{\partial{T}}\approx-\frac{1}{2T}<0.   \label{eq41}
\end{equation}
A more detailed analysis of the Casimir entropy within the Lifshitz framework gives:
\begin{gather}
S_L(a,T)=-\frac{1}{2}\log\left[\frac{2\pi\mathsf{T}}{\Lambda}(d+2)\right]-\frac{1}{2} \notag \\
-\sum_{n=1}^{\infty}\log\left[1-\frac{e^{-4\pi{n}\mathsf{T}d}}{(1+4\pi{n}\mathsf{T})^2}\right] \notag \\
-\sum_{n=1}^{\infty}\frac{4\pi{n}\mathsf{T}(4\pi{n}\mathsf{T}d+d+2)}{(4\pi{n}\mathsf{T}+1)\left[(4\pi{n}\mathsf{T}+1)^2e^{4\pi{n}\mathsf{T}d}-1\right]}.  \label{eq42}
\end{gather}
This formulation leads to a dilemma: if the zero-frequency term (the first line in Eq. (\ref{eq42})) is retained, the entropy at zero temperature does not vanish but diverges to positive infinity, violating the third law of thermodynamics. If this term is omitted, the finite-temperature Casimir entropy (the summations in Eq. (\ref{eq42})) becomes negative, which also contradicts the third law. In contrast, the canonical quantization method avoids these inconsistencies. As the temperature approaches zero, the wave-vector integral is suppressed by the factor $\left(\frac{q}{2\mathsf{T}}\right)^2\text{csch}^2\left(\frac{q}{2\mathsf{T}}\right)$ and tends to zero in the limit $\mathsf{T}\rightarrow0$, ensuring compliance with the third law of thermodynamics.

We now turn to the issue of infrared divergence. The observed logarithmic infrared divergence has a clear physical origin. Examining the eigenmodes in the long-wavelength limit ($k\rightarrow0$), we find 
\begin{gather}
B_{k}\rightarrow{-1},~~~  C_{k}\rightarrow\frac{v^2}{\gamma{a}+2v^2}, \label{eq43} \\
  D_{k}\rightarrow-\frac{v^2}{\gamma{a}+2v^2},~~~  G_{k}\rightarrow0,   \label{eq44}
\end{gather}
which differs markedly from the behavior in the ultraviolet regime ($|k|\rightarrow\infty$), where $B_{k}\rightarrow0$, $C_{k}\rightarrow1$, $D_{k}\rightarrow0$, and $G_{k}\rightarrow1$. 
This contrast reveals that while high-energy modes remain essentially unaffected by the barriers and do not contribute to the Casimir interaction—eliminating the need for ultraviolet counterterms—the low-energy, long-wavelength modes do participate significantly. Although their low energy makes their contribution to the Casimir force negligible, these modes become particularly important at finite temperatures. Due to their higher thermodynamic occupancy and enhanced fluctuations, they dominate the entropy accumulation at large distances, thereby giving rise to the observed infrared divergence in the Casimir entropy.

\section{Summary}  \label{sec5}
In this work, we have examined the finite-temperature Casimir effect in a (1+1)-dimensional scalar field system in the presence of two delta-function potentials. Using the canonical quantization approach, we have derived explicit expressions for both the Casimir force and the Casimir entropy, and carried out a systematic comparison with corresponding results obtained via the Lifshitz formula.
At zero temperature, high-precision numerical evaluations confirm that the Casimir forces calculated through both formalisms coincide exactly, thereby validating the internal consistency of the canonical quantization framework in this regime. However, notable discrepancies emerge at finite temperatures. In particular, in the large-separation limit, the Casimir force obtained from canonical quantization is precisely one-half the magnitude predicted by the Lifshitz theory. More significantly, the entropy derived within the canonical quantization approach remains fully consistent with the third law of thermodynamics---vanishing as the temperature approaches zero. In contrast, the Lifshitz formulation leads to thermodynamic inconsistencies, manifesting either as a divergence of entropy at zero temperature or the appearance of negative entropy values at finite temperatures.

\section{Acknowledgments}
This research was supported by the National Natural Science Foundation of China under Grant No. 12174101.

\bibliography{references.bib}

\begin{thebibliography}{33}%
\makeatletter
\providecommand \@ifxundefined [1]{%
 \@ifx{#1\undefined}
}%
\providecommand \@ifnum [1]{%
 \ifnum #1\expandafter \@firstoftwo
 \else \expandafter \@secondoftwo
 \fi
}%
\providecommand \@ifx [1]{%
 \ifx #1\expandafter \@firstoftwo
 \else \expandafter \@secondoftwo
 \fi
}%
\providecommand \natexlab [1]{#1}%
\providecommand \enquote  [1]{``#1''}%
\providecommand \bibnamefont  [1]{#1}%
\providecommand \bibfnamefont [1]{#1}%
\providecommand \citenamefont [1]{#1}%
\providecommand \href@noop [0]{\@secondoftwo}%
\providecommand \href [0]{\begingroup \@sanitize@url \@href}%
\providecommand \@href[1]{\@@startlink{#1}\@@href}%
\providecommand \@@href[1]{\endgroup#1\@@endlink}%
\providecommand \@sanitize@url [0]{\catcode `\\12\catcode `\$12\catcode
  `\&12\catcode `\#12\catcode `\^12\catcode `\_12\catcode `\%12\relax}%
\providecommand \@@startlink[1]{}%
\providecommand \@@endlink[0]{}%
\providecommand \url  [0]{\begingroup\@sanitize@url \@url }%
\providecommand \@url [1]{\endgroup\@href {#1}{\urlprefix }}%
\providecommand \urlprefix  [0]{URL }%
\providecommand \Eprint [0]{\href }%
\providecommand \doibase [0]{https://doi.org/}%
\providecommand \selectlanguage [0]{\@gobble}%
\providecommand \bibinfo  [0]{\@secondoftwo}%
\providecommand \bibfield  [0]{\@secondoftwo}%
\providecommand \translation [1]{[#1]}%
\providecommand \BibitemOpen [0]{}%
\providecommand \bibitemStop [0]{}%
\providecommand \bibitemNoStop [0]{.\EOS\space}%
\providecommand \EOS [0]{\spacefactor3000\relax}%
\providecommand \BibitemShut  [1]{\csname bibitem#1\endcsname}%
\let\auto@bib@innerbib\@empty
\bibitem [{\citenamefont {Casimir}(1948)}]{Casimir1948PKNAW}%
  \BibitemOpen
  \bibfield  {author} {\bibinfo {author} {\bibfnamefont {H.~B.~G.}\
  \bibnamefont {Casimir}},\ }\bibfield  {title} {\bibinfo {title} {On the
  attraction between two perfectly conducting plates},\ }\href@noop {}
  {\bibfield  {journal} {\bibinfo  {journal} {Proc. Kon. Nederland. Akad.
  Wetensch.}\ }\textbf {\bibinfo {volume} {51}},\ \bibinfo {pages} {793}
  (\bibinfo {year} {1948})}\BibitemShut {NoStop}%
\bibitem [{\citenamefont {Milonni}(1994)}]{Milonni1994QVacuumbook}%
  \BibitemOpen
  \bibfield  {author} {\bibinfo {author} {\bibfnamefont {P.~W.}\ \bibnamefont
  {Milonni}},\ }\href@noop {} {\emph {\bibinfo {title} {The Quantum Vacuum: An
  Introduction to Quantum Electrodynamics}}}\ (\bibinfo  {publisher} {Academic
  Press, San Diego},\ \bibinfo {year} {1994})\BibitemShut {NoStop}%
\bibitem [{\citenamefont {Mostepanenko}\ \emph {et~al.}(1997)\citenamefont
  {Mostepanenko}, \citenamefont {Trunov},\ and\ \citenamefont
  {Znajek}}]{Mostepanenko1997Casimirbook}%
  \BibitemOpen
  \bibfield  {author} {\bibinfo {author} {\bibfnamefont {V.~M.}\ \bibnamefont
  {Mostepanenko}}, \bibinfo {author} {\bibfnamefont {N.~N.}\ \bibnamefont
  {Trunov}},\ and\ \bibinfo {author} {\bibfnamefont {R.~L.}\ \bibnamefont
  {Znajek}},\ }\href@noop {} {\emph {\bibinfo {title} {The Casimir Effect and
  its Applications}}}\ (\bibinfo  {publisher} {Oxford University Press,
  Oxford},\ \bibinfo {year} {1997})\BibitemShut {NoStop}%
\bibitem [{\citenamefont {Bordag}\ \emph {et~al.}(2009)\citenamefont {Bordag},
  \citenamefont {Klimchitskaya}, \citenamefont {Mohideen},\ and\ \citenamefont
  {Mostepanenko}}]{Bordag2009Casimirbook}%
  \BibitemOpen
  \bibfield  {author} {\bibinfo {author} {\bibfnamefont {M.}~\bibnamefont
  {Bordag}}, \bibinfo {author} {\bibfnamefont {G.~L.}\ \bibnamefont
  {Klimchitskaya}}, \bibinfo {author} {\bibfnamefont {U.}~\bibnamefont
  {Mohideen}},\ and\ \bibinfo {author} {\bibfnamefont {V.~M.}\ \bibnamefont
  {Mostepanenko}},\ }\href@noop {} {\emph {\bibinfo {title} {Advances in the
  Casimir Effect}}}\ (\bibinfo  {publisher} {Oxford University Press, Oxford},\
  \bibinfo {year} {2009})\BibitemShut {NoStop}%
\bibitem [{\citenamefont {Bordag}\ \emph {et~al.}(2001)\citenamefont {Bordag},
  \citenamefont {Mohideen},\ and\ \citenamefont
  {Mostepanenko}}]{Bordag2001PhysRep}%
  \BibitemOpen
  \bibfield  {author} {\bibinfo {author} {\bibfnamefont {M.}~\bibnamefont
  {Bordag}}, \bibinfo {author} {\bibfnamefont {U.}~\bibnamefont {Mohideen}},\
  and\ \bibinfo {author} {\bibfnamefont {V.~M.}\ \bibnamefont {Mostepanenko}},\
  }\bibfield  {title} {\bibinfo {title} {New developments in the casimir
  effect},\ }\href
  {https://doi.org/https://doi.org/10.1016/S0370-1573(01)00015-1} {\bibfield
  {journal} {\bibinfo  {journal} {Physics Reports}\ }\textbf {\bibinfo {volume}
  {353}},\ \bibinfo {pages} {1} (\bibinfo {year} {2001})}\BibitemShut {NoStop}%
\bibitem [{\citenamefont {Buks}\ and\ \citenamefont
  {Roukes}(2001)}]{BuksE2001PRB}%
  \BibitemOpen
  \bibfield  {author} {\bibinfo {author} {\bibfnamefont {E.}~\bibnamefont
  {Buks}}\ and\ \bibinfo {author} {\bibfnamefont {M.~L.}\ \bibnamefont
  {Roukes}},\ }\bibfield  {title} {\bibinfo {title} {Stiction, adhesion energy,
  and the casimir effect in micromechanical systems},\ }\href
  {https://doi.org/10.1103/PhysRevB.63.033402} {\bibfield  {journal} {\bibinfo
  {journal} {Phys. Rev. B}\ }\textbf {\bibinfo {volume} {63}},\ \bibinfo
  {pages} {033402} (\bibinfo {year} {2001})}\BibitemShut {NoStop}%
\bibitem [{\citenamefont {Serry}\ \emph {et~al.}(1998)\citenamefont {Serry},
  \citenamefont {Walliser},\ and\ \citenamefont {Maclay}}]{SerryFM1998JAP}%
  \BibitemOpen
  \bibfield  {author} {\bibinfo {author} {\bibfnamefont {F.~M.}\ \bibnamefont
  {Serry}}, \bibinfo {author} {\bibfnamefont {D.}~\bibnamefont {Walliser}},\
  and\ \bibinfo {author} {\bibfnamefont {G.~J.}\ \bibnamefont {Maclay}},\
  }\bibfield  {title} {\bibinfo {title} {The role of the casimir effect in the
  static deflection and stiction of membrane strips in microelectromechanical
  systems (mems)},\ }\href {https://doi.org/10.1063/1.368410} {\bibfield
  {journal} {\bibinfo  {journal} {Journal of Applied Physics}\ }\textbf
  {\bibinfo {volume} {84}},\ \bibinfo {pages} {2501} (\bibinfo {year}
  {1998})}\BibitemShut {NoStop}%
\bibitem [{\citenamefont {Rodriguez}\ \emph {et~al.}(2011)\citenamefont
  {Rodriguez}, \citenamefont {Woolf}, \citenamefont {Hui}, \citenamefont
  {Iwase}, \citenamefont {McCauley}, \citenamefont {Capasso}, \citenamefont
  {Loncar},\ and\ \citenamefont {Johnson}}]{RodriguezAW2011APL}%
  \BibitemOpen
  \bibfield  {author} {\bibinfo {author} {\bibfnamefont {A.~W.}\ \bibnamefont
  {Rodriguez}}, \bibinfo {author} {\bibfnamefont {D.}~\bibnamefont {Woolf}},
  \bibinfo {author} {\bibfnamefont {P.-C.}\ \bibnamefont {Hui}}, \bibinfo
  {author} {\bibfnamefont {E.}~\bibnamefont {Iwase}}, \bibinfo {author}
  {\bibfnamefont {A.~P.}\ \bibnamefont {McCauley}}, \bibinfo {author}
  {\bibfnamefont {F.}~\bibnamefont {Capasso}}, \bibinfo {author} {\bibfnamefont
  {M.}~\bibnamefont {Loncar}},\ and\ \bibinfo {author} {\bibfnamefont {S.~G.}\
  \bibnamefont {Johnson}},\ }\bibfield  {title} {\bibinfo {title} {Designing
  evanescent optical interactions to control the expression of casimir forces
  in optomechanical structures},\ }\href {https://doi.org/10.1063/1.3589119}
  {\bibfield  {journal} {\bibinfo  {journal} {Applied Physics Letters}\
  }\textbf {\bibinfo {volume} {98}},\ \bibinfo {pages} {194105} (\bibinfo
  {year} {2011})}\BibitemShut {NoStop}%
\bibitem [{\citenamefont {Sushkov}\ \emph {et~al.}(2011)\citenamefont
  {Sushkov}, \citenamefont {Kim}, \citenamefont {Dalvit},\ and\ \citenamefont
  {Lamoreaux}}]{SushkovAO2011NatPhys}%
  \BibitemOpen
  \bibfield  {author} {\bibinfo {author} {\bibfnamefont {A.~O.}\ \bibnamefont
  {Sushkov}}, \bibinfo {author} {\bibfnamefont {W.~J.}\ \bibnamefont {Kim}},
  \bibinfo {author} {\bibfnamefont {D.~A.~R.}\ \bibnamefont {Dalvit}},\ and\
  \bibinfo {author} {\bibfnamefont {S.~K.}\ \bibnamefont {Lamoreaux}},\
  }\bibfield  {title} {\bibinfo {title} {{Observation of the thermal Casimir
  force}},\ }\href {https://doi.org/10.1038/nphys1909} {\bibfield  {journal}
  {\bibinfo  {journal} {Nat. Phys.}\ }\textbf {\bibinfo {volume} {7}},\
  \bibinfo {pages} {230} (\bibinfo {year} {2011})}\BibitemShut {NoStop}%
\bibitem [{\citenamefont {Mostepanenko}\ \emph {et~al.}(2006)\citenamefont
  {Mostepanenko}, \citenamefont {Bezerra}, \citenamefont {Decca}, \citenamefont
  {Geyer}, \citenamefont {Fischbach}, \citenamefont {Klimchitskaya},
  \citenamefont {Krause}, \citenamefont {L{\'{o}}pez},\ and\ \citenamefont
  {Romero}}]{Mostepanenko2006JPAMG}%
  \BibitemOpen
  \bibfield  {author} {\bibinfo {author} {\bibfnamefont {V.~M.}\ \bibnamefont
  {Mostepanenko}}, \bibinfo {author} {\bibfnamefont {V.~B.}\ \bibnamefont
  {Bezerra}}, \bibinfo {author} {\bibfnamefont {R.~S.}\ \bibnamefont {Decca}},
  \bibinfo {author} {\bibfnamefont {B.}~\bibnamefont {Geyer}}, \bibinfo
  {author} {\bibfnamefont {E.}~\bibnamefont {Fischbach}}, \bibinfo {author}
  {\bibfnamefont {G.~L.}\ \bibnamefont {Klimchitskaya}}, \bibinfo {author}
  {\bibfnamefont {D.~E.}\ \bibnamefont {Krause}}, \bibinfo {author}
  {\bibfnamefont {D.}~\bibnamefont {L{\'{o}}pez}},\ and\ \bibinfo {author}
  {\bibfnamefont {C.}~\bibnamefont {Romero}},\ }\bibfield  {title} {\bibinfo
  {title} {Present status of controversies regarding the thermal casimir
  force},\ }\href {https://doi.org/10.1088/0305-4470/39/21/S58} {\bibfield
  {journal} {\bibinfo  {journal} {J. Phys. A Math. Gen.}\ }\textbf {\bibinfo
  {volume} {39}},\ \bibinfo {pages} {6589} (\bibinfo {year}
  {2006})}\BibitemShut {NoStop}%
\bibitem [{\citenamefont {Mostepanenko}\ and\ \citenamefont
  {Geyer}(2008)}]{Mostepanenko2008JPA}%
  \BibitemOpen
  \bibfield  {author} {\bibinfo {author} {\bibfnamefont {V.~M.}\ \bibnamefont
  {Mostepanenko}}\ and\ \bibinfo {author} {\bibfnamefont {B.}~\bibnamefont
  {Geyer}},\ }\bibfield  {title} {\bibinfo {title} {New approach to the thermal
  casimir force between real metals},\ }\href
  {https://doi.org/10.1088/1751-8113/41/16/164014} {\bibfield  {journal}
  {\bibinfo  {journal} {J. PHYS. A-MATH. THEOR.}\ }\textbf {\bibinfo {volume}
  {41}},\ \bibinfo {pages} {164014} (\bibinfo {year} {2008})}\BibitemShut
  {NoStop}%
\bibitem [{\citenamefont {Klimchitskaya}\ and\ \citenamefont
  {Geyer}(2008)}]{Klimchitskaya2008JPAMT}%
  \BibitemOpen
  \bibfield  {author} {\bibinfo {author} {\bibfnamefont {G.~L.}\ \bibnamefont
  {Klimchitskaya}}\ and\ \bibinfo {author} {\bibfnamefont {B.}~\bibnamefont
  {Geyer}},\ }\bibfield  {title} {\bibinfo {title} {Problems in the theory of
  the thermal casimir force between dielectrics and semiconductors},\ }\href
  {https://doi.org/10.1088/1751-8113/41/16/164032} {\bibfield  {journal}
  {\bibinfo  {journal} {J. PHYS. A-MATH. THEOR.}\ }\textbf {\bibinfo {volume}
  {41}},\ \bibinfo {pages} {164032} (\bibinfo {year} {2008})}\BibitemShut
  {NoStop}%
\bibitem [{\citenamefont {Lamoreaux}(2012)}]{Lamoreaux2012ARNPS}%
  \BibitemOpen
  \bibfield  {author} {\bibinfo {author} {\bibfnamefont {S.~K.}\ \bibnamefont
  {Lamoreaux}},\ }\bibfield  {title} {\bibinfo {title} {The casimir force and
  related effects: The status of the finite temperature correction and limits
  on new long-range forces},\ }\href
  {https://doi.org/https://doi.org/10.1146/annurev-nucl-102711-095013}
  {\bibfield  {journal} {\bibinfo  {journal} {Annu. Rev. Nucl. Part. S.}\
  }\textbf {\bibinfo {volume} {62}},\ \bibinfo {pages} {37} (\bibinfo {year}
  {2012})}\BibitemShut {NoStop}%
\bibitem [{\citenamefont {Hartmann}\ \emph {et~al.}(2017)\citenamefont
  {Hartmann}, \citenamefont {Ingold},\ and\ \citenamefont
  {Neto}}]{HartmannM2017PRL}%
  \BibitemOpen
  \bibfield  {author} {\bibinfo {author} {\bibfnamefont {M.}~\bibnamefont
  {Hartmann}}, \bibinfo {author} {\bibfnamefont {G.-L.}\ \bibnamefont
  {Ingold}},\ and\ \bibinfo {author} {\bibfnamefont {P.~A.~M.}\ \bibnamefont
  {Neto}},\ }\bibfield  {title} {\bibinfo {title} {Plasma versus drude modeling
  of the casimir force: Beyond the proximity force approximation},\ }\href
  {https://doi.org/10.1103/PhysRevLett.119.043901} {\bibfield  {journal}
  {\bibinfo  {journal} {Phys. Rev. Lett.}\ }\textbf {\bibinfo {volume} {119}},\
  \bibinfo {pages} {043901} (\bibinfo {year} {2017})}\BibitemShut {NoStop}%
\bibitem [{\citenamefont {Liu}\ \emph {et~al.}(2019)\citenamefont {Liu},
  \citenamefont {Xu}, \citenamefont {Klimchitskaya}, \citenamefont
  {Mostepanenko},\ and\ \citenamefont {Mohideen}}]{LiuM2019PRB}%
  \BibitemOpen
  \bibfield  {author} {\bibinfo {author} {\bibfnamefont {M.}~\bibnamefont
  {Liu}}, \bibinfo {author} {\bibfnamefont {J.}~\bibnamefont {Xu}}, \bibinfo
  {author} {\bibfnamefont {G.~L.}\ \bibnamefont {Klimchitskaya}}, \bibinfo
  {author} {\bibfnamefont {V.~M.}\ \bibnamefont {Mostepanenko}},\ and\ \bibinfo
  {author} {\bibfnamefont {U.}~\bibnamefont {Mohideen}},\ }\bibfield  {title}
  {\bibinfo {title} {Examining the casimir puzzle with an upgraded afm-based
  technique and advanced surface cleaning},\ }\href
  {https://doi.org/10.1103/PhysRevB.100.081406} {\bibfield  {journal} {\bibinfo
   {journal} {Phys. Rev. B}\ }\textbf {\bibinfo {volume} {100}},\ \bibinfo
  {pages} {081406} (\bibinfo {year} {2019})}\BibitemShut {NoStop}%
\bibitem [{\citenamefont {Klimchitskaya}\ and\ \citenamefont
  {Mostepanenko}(2022)}]{Klimchitskaya2022IJMPA}%
  \BibitemOpen
  \bibfield  {author} {\bibinfo {author} {\bibfnamefont {G.~L.}\ \bibnamefont
  {Klimchitskaya}}\ and\ \bibinfo {author} {\bibfnamefont {V.~M.}\ \bibnamefont
  {Mostepanenko}},\ }\bibfield  {title} {\bibinfo {title} {Current status of
  the problem of thermal casimir force},\ }\href
  {https://doi.org/10.1142/S0217751X22410020} {\bibfield  {journal} {\bibinfo
  {journal} {INT. J. MOD. PHYS. A}\ }\textbf {\bibinfo {volume} {37}},\
  \bibinfo {pages} {2241002} (\bibinfo {year} {2022})}\BibitemShut {NoStop}%
\bibitem [{\citenamefont {Mostepanenko}(2021)}]{Mostepanenko2021Universe}%
  \BibitemOpen
  \bibfield  {author} {\bibinfo {author} {\bibfnamefont {V.~M.}\ \bibnamefont
  {Mostepanenko}},\ }\bibfield  {title} {\bibinfo {title} {Casimir puzzle and
  casimir conundrum: Discovery and search for resolution},\ }\bibfield
  {journal} {\bibinfo  {journal} {Universe}\ }\textbf {\bibinfo {volume} {7}},\
  \href {https://doi.org/10.3390/universe7040084} {10.3390/universe7040084}
  (\bibinfo {year} {2021})\BibitemShut {NoStop}%
\bibitem [{\citenamefont {Rodriguez-Lopez}(2011)}]{RodriguezLopezP2011PRB}%
  \BibitemOpen
  \bibfield  {author} {\bibinfo {author} {\bibfnamefont {P.}~\bibnamefont
  {Rodriguez-Lopez}},\ }\bibfield  {title} {\bibinfo {title} {Casimir energy
  and entropy in the sphere-sphere geometry},\ }\href
  {https://doi.org/10.1103/PhysRevB.84.075431} {\bibfield  {journal} {\bibinfo
  {journal} {Phys. Rev. B}\ }\textbf {\bibinfo {volume} {84}},\ \bibinfo
  {pages} {075431} (\bibinfo {year} {2011})}\BibitemShut {NoStop}%
\bibitem [{\citenamefont {Ingold}\ \emph {et~al.}(2015)\citenamefont {Ingold},
  \citenamefont {Umrath}, \citenamefont {Hartmann}, \citenamefont {Gu\'erout},
  \citenamefont {Lambrecht}, \citenamefont {Reynaud},\ and\ \citenamefont
  {Milton}}]{IngoldGL2015PRE}%
  \BibitemOpen
  \bibfield  {author} {\bibinfo {author} {\bibfnamefont {G.-L.}\ \bibnamefont
  {Ingold}}, \bibinfo {author} {\bibfnamefont {S.}~\bibnamefont {Umrath}},
  \bibinfo {author} {\bibfnamefont {M.}~\bibnamefont {Hartmann}}, \bibinfo
  {author} {\bibfnamefont {R.}~\bibnamefont {Gu\'erout}}, \bibinfo {author}
  {\bibfnamefont {A.}~\bibnamefont {Lambrecht}}, \bibinfo {author}
  {\bibfnamefont {S.}~\bibnamefont {Reynaud}},\ and\ \bibinfo {author}
  {\bibfnamefont {K.~A.}\ \bibnamefont {Milton}},\ }\bibfield  {title}
  {\bibinfo {title} {Geometric origin of negative casimir entropies: A
  scattering-channel analysis},\ }\href
  {https://doi.org/10.1103/PhysRevE.91.033203} {\bibfield  {journal} {\bibinfo
  {journal} {Phys. Rev. E}\ }\textbf {\bibinfo {volume} {91}},\ \bibinfo
  {pages} {033203} (\bibinfo {year} {2015})}\BibitemShut {NoStop}%
\bibitem [{\citenamefont {Milton}\ \emph {et~al.}(2017)\citenamefont {Milton},
  \citenamefont {Li}, \citenamefont {Kalauni}, \citenamefont {Parashar},
  \citenamefont {Gu\'{e}rout}, \citenamefont {Ingold}, \citenamefont
  {Lambrecht},\ and\ \citenamefont {Reynaud}}]{MiltonKA2016FdP}%
  \BibitemOpen
  \bibfield  {author} {\bibinfo {author} {\bibfnamefont {K.~A.}\ \bibnamefont
  {Milton}}, \bibinfo {author} {\bibfnamefont {Y.}~\bibnamefont {Li}}, \bibinfo
  {author} {\bibfnamefont {P.}~\bibnamefont {Kalauni}}, \bibinfo {author}
  {\bibfnamefont {P.}~\bibnamefont {Parashar}}, \bibinfo {author}
  {\bibfnamefont {R.}~\bibnamefont {Gu\'{e}rout}}, \bibinfo {author}
  {\bibfnamefont {G.-L.}\ \bibnamefont {Ingold}}, \bibinfo {author}
  {\bibfnamefont {A.}~\bibnamefont {Lambrecht}},\ and\ \bibinfo {author}
  {\bibfnamefont {S.}~\bibnamefont {Reynaud}},\ }\bibfield  {title} {\bibinfo
  {title} {Negative entropies in casimir and casimir-polder interactions},\
  }\href {https://doi.org/https://doi.org/10.1002/prop.201600047} {\bibfield
  {journal} {\bibinfo  {journal} {Fortschritte der Physik}\ }\textbf {\bibinfo
  {volume} {65}},\ \bibinfo {pages} {1600047} (\bibinfo {year}
  {2017})}\BibitemShut {NoStop}%
\bibitem [{\citenamefont {Bordag}\ and\ \citenamefont
  {Kirsten}(2018)}]{BordagM2018JPA}%
  \BibitemOpen
  \bibfield  {author} {\bibinfo {author} {\bibfnamefont {M.}~\bibnamefont
  {Bordag}}\ and\ \bibinfo {author} {\bibfnamefont {K.}~\bibnamefont
  {Kirsten}},\ }\bibfield  {title} {\bibinfo {title} {On the entropy of a
  spherical plasma shell},\ }\href {https://doi.org/10.1088/1751-8121/aae4c1}
  {\bibfield  {journal} {\bibinfo  {journal} {Journal of Physics A:
  Mathematical and Theoretical}\ }\textbf {\bibinfo {volume} {51}},\ \bibinfo
  {pages} {455001} (\bibinfo {year} {2018})}\BibitemShut {NoStop}%
\bibitem [{\citenamefont {Li}\ \emph {et~al.}(2016)\citenamefont {Li},
  \citenamefont {Milton}, \citenamefont {Kalauni},\ and\ \citenamefont
  {Parashar}}]{LiY2016PRD}%
  \BibitemOpen
  \bibfield  {author} {\bibinfo {author} {\bibfnamefont {Y.}~\bibnamefont
  {Li}}, \bibinfo {author} {\bibfnamefont {K.~A.}\ \bibnamefont {Milton}},
  \bibinfo {author} {\bibfnamefont {P.}~\bibnamefont {Kalauni}},\ and\ \bibinfo
  {author} {\bibfnamefont {P.}~\bibnamefont {Parashar}},\ }\bibfield  {title}
  {\bibinfo {title} {Casimir self-entropy of an electromagnetic thin sheet},\
  }\href {https://doi.org/10.1103/PhysRevD.94.085010} {\bibfield  {journal}
  {\bibinfo  {journal} {Phys. Rev. D}\ }\textbf {\bibinfo {volume} {94}},\
  \bibinfo {pages} {085010} (\bibinfo {year} {2016})}\BibitemShut {NoStop}%
\bibitem [{\citenamefont {Bordag}(2018)}]{BordagM2018PRD}%
  \BibitemOpen
  \bibfield  {author} {\bibinfo {author} {\bibfnamefont {M.}~\bibnamefont
  {Bordag}},\ }\bibfield  {title} {\bibinfo {title} {Free energy and entropy
  for thin sheets},\ }\href {https://doi.org/10.1103/PhysRevD.98.085010}
  {\bibfield  {journal} {\bibinfo  {journal} {Phys. Rev. D}\ }\textbf {\bibinfo
  {volume} {98}},\ \bibinfo {pages} {085010} (\bibinfo {year}
  {2018})}\BibitemShut {NoStop}%
\bibitem [{\citenamefont {Bordag}\ \emph {et~al.}(2020)\citenamefont {Bordag},
  \citenamefont {Mu{\~{n}}oz-Casta{\~{n}}eda},\ and\ \citenamefont
  {Santamar{\'i}a-Sanz}}]{BordagM2020EPJC}%
  \BibitemOpen
  \bibfield  {author} {\bibinfo {author} {\bibfnamefont {M.}~\bibnamefont
  {Bordag}}, \bibinfo {author} {\bibfnamefont {J.~M.}\ \bibnamefont
  {Mu{\~{n}}oz-Casta{\~{n}}eda}},\ and\ \bibinfo {author} {\bibfnamefont
  {L.}~\bibnamefont {Santamar{\'i}a-Sanz}},\ }\bibfield  {title} {\bibinfo
  {title} {Free energy and entropy for finite temperature quantum field theory
  under the influence of periodic backgrounds},\ }\href
  {https://doi.org/10.1140/epjc/s10052-020-7783-3} {\bibfield  {journal}
  {\bibinfo  {journal} {The European Physical Journal C}\ }\textbf {\bibinfo
  {volume} {80}},\ \bibinfo {pages} {221} (\bibinfo {year} {2020})}\BibitemShut
  {NoStop}%
\bibitem [{\citenamefont {Bordag}\ \emph {et~al.}(1992)\citenamefont {Bordag},
  \citenamefont {Hennig},\ and\ \citenamefont {Robaschik}}]{BordagM1992JPA}%
  \BibitemOpen
  \bibfield  {author} {\bibinfo {author} {\bibfnamefont {M.}~\bibnamefont
  {Bordag}}, \bibinfo {author} {\bibfnamefont {D.}~\bibnamefont {Hennig}},\
  and\ \bibinfo {author} {\bibfnamefont {D.}~\bibnamefont {Robaschik}},\
  }\bibfield  {title} {\bibinfo {title} {Vacuum energy in quantum field theory
  with external potentials concentrated on planes},\ }\href
  {https://doi.org/10.1088/0305-4470/25/16/023} {\bibfield  {journal} {\bibinfo
   {journal} {Journal of Physics A: Mathematical and General}\ }\textbf
  {\bibinfo {volume} {25}},\ \bibinfo {pages} {4483} (\bibinfo {year}
  {1992})}\BibitemShut {NoStop}%
\bibitem [{\citenamefont {Graham}\ \emph {et~al.}(2002)\citenamefont {Graham},
  \citenamefont {Jaffe}, \citenamefont {Khemani}, \citenamefont {Quandt},
  \citenamefont {Scandurra},\ and\ \citenamefont {Weigel}}]{GrahamN2002NPB}%
  \BibitemOpen
  \bibfield  {author} {\bibinfo {author} {\bibfnamefont {N.}~\bibnamefont
  {Graham}}, \bibinfo {author} {\bibfnamefont {R.}~\bibnamefont {Jaffe}},
  \bibinfo {author} {\bibfnamefont {V.}~\bibnamefont {Khemani}}, \bibinfo
  {author} {\bibfnamefont {M.}~\bibnamefont {Quandt}}, \bibinfo {author}
  {\bibfnamefont {M.}~\bibnamefont {Scandurra}},\ and\ \bibinfo {author}
  {\bibfnamefont {H.}~\bibnamefont {Weigel}},\ }\bibfield  {title} {\bibinfo
  {title} {Calculating vacuum energies in renormalizable quantum field
  theories:: A new approach to the casimir problem},\ }\href
  {https://doi.org/https://doi.org/10.1016/S0550-3213(02)00823-4} {\bibfield
  {journal} {\bibinfo  {journal} {Nuclear Physics B}\ }\textbf {\bibinfo
  {volume} {645}},\ \bibinfo {pages} {49} (\bibinfo {year} {2002})}\BibitemShut
  {NoStop}%
\bibitem [{\citenamefont {Graham}\ \emph {et~al.}(2003)\citenamefont {Graham},
  \citenamefont {Jaffe}, \citenamefont {Khemani}, \citenamefont {Quandt},
  \citenamefont {Scandurra},\ and\ \citenamefont {Weigel}}]{GrahamN2003PLB}%
  \BibitemOpen
  \bibfield  {author} {\bibinfo {author} {\bibfnamefont {N.}~\bibnamefont
  {Graham}}, \bibinfo {author} {\bibfnamefont {R.}~\bibnamefont {Jaffe}},
  \bibinfo {author} {\bibfnamefont {V.}~\bibnamefont {Khemani}}, \bibinfo
  {author} {\bibfnamefont {M.}~\bibnamefont {Quandt}}, \bibinfo {author}
  {\bibfnamefont {M.}~\bibnamefont {Scandurra}},\ and\ \bibinfo {author}
  {\bibfnamefont {H.}~\bibnamefont {Weigel}},\ }\bibfield  {title} {\bibinfo
  {title} {Casimir energies in light of quantum field theory},\ }\href
  {https://doi.org/https://doi.org/10.1016/j.physletb.2003.03.003} {\bibfield
  {journal} {\bibinfo  {journal} {Physics Letters B}\ }\textbf {\bibinfo
  {volume} {572}},\ \bibinfo {pages} {196} (\bibinfo {year}
  {2003})}\BibitemShut {NoStop}%
\bibitem [{\citenamefont {Graham}\ \emph {et~al.}(2004)\citenamefont {Graham},
  \citenamefont {Jaffe}, \citenamefont {Khemani}, \citenamefont {Quandt},
  \citenamefont {Schr\"{o}der},\ and\ \citenamefont {Weigel}}]{GrahamN2004NPB}%
  \BibitemOpen
  \bibfield  {author} {\bibinfo {author} {\bibfnamefont {N.}~\bibnamefont
  {Graham}}, \bibinfo {author} {\bibfnamefont {R.}~\bibnamefont {Jaffe}},
  \bibinfo {author} {\bibfnamefont {V.}~\bibnamefont {Khemani}}, \bibinfo
  {author} {\bibfnamefont {M.}~\bibnamefont {Quandt}}, \bibinfo {author}
  {\bibfnamefont {O.}~\bibnamefont {Schr\"{o}der}},\ and\ \bibinfo {author}
  {\bibfnamefont {H.}~\bibnamefont {Weigel}},\ }\bibfield  {title} {\bibinfo
  {title} {The dirichlet casimir problem},\ }\href
  {https://doi.org/https://doi.org/10.1016/j.nuclphysb.2003.11.001} {\bibfield
  {journal} {\bibinfo  {journal} {Nuclear Physics B}\ }\textbf {\bibinfo
  {volume} {677}},\ \bibinfo {pages} {379} (\bibinfo {year}
  {2004})}\BibitemShut {NoStop}%
\bibitem [{\citenamefont {Milton}(2004)}]{MiltonKA2004JPA}%
  \BibitemOpen
  \bibfield  {author} {\bibinfo {author} {\bibfnamefont {K.~A.}\ \bibnamefont
  {Milton}},\ }\bibfield  {title} {\bibinfo {title} {Casimir energies and
  pressures for $\delta$-function potentials},\ }\href
  {https://doi.org/10.1088/0305-4470/37/24/014} {\bibfield  {journal} {\bibinfo
   {journal} {Journal of Physics A: Mathematical and General}\ }\textbf
  {\bibinfo {volume} {37}},\ \bibinfo {pages} {6391} (\bibinfo {year}
  {2004})}\BibitemShut {NoStop}%
\bibitem [{\citenamefont {Kupiszewska}\ and\ \citenamefont
  {Mostowski}(1990)}]{KupiszewskaD1990PRA}%
  \BibitemOpen
  \bibfield  {author} {\bibinfo {author} {\bibfnamefont {D.}~\bibnamefont
  {Kupiszewska}}\ and\ \bibinfo {author} {\bibfnamefont {J.}~\bibnamefont
  {Mostowski}},\ }\bibfield  {title} {\bibinfo {title} {Casimir effect for
  dielectric plates},\ }\href {https://doi.org/10.1103/PhysRevA.41.4636}
  {\bibfield  {journal} {\bibinfo  {journal} {Phys. Rev. A}\ }\textbf {\bibinfo
  {volume} {41}},\ \bibinfo {pages} {4636} (\bibinfo {year}
  {1990})}\BibitemShut {NoStop}%
\bibitem [{\citenamefont {Kupiszewska}(1992)}]{KupiszewskaD1992PRA}%
  \BibitemOpen
  \bibfield  {author} {\bibinfo {author} {\bibfnamefont {D.}~\bibnamefont
  {Kupiszewska}},\ }\bibfield  {title} {\bibinfo {title} {Casimir effect in
  absorbing media},\ }\href {https://doi.org/10.1103/PhysRevA.46.2286}
  {\bibfield  {journal} {\bibinfo  {journal} {Phys. Rev. A}\ }\textbf {\bibinfo
  {volume} {46}},\ \bibinfo {pages} {2286} (\bibinfo {year}
  {1992})}\BibitemShut {NoStop}%
\bibitem [{\citenamefont {van Enk}(1995)}]{vanEnkSJ1995PRA}%
  \BibitemOpen
  \bibfield  {author} {\bibinfo {author} {\bibfnamefont {S.~J.}\ \bibnamefont
  {van Enk}},\ }\bibfield  {title} {\bibinfo {title} {Casimir torque between
  dielectrics},\ }\href {https://doi.org/10.1103/PhysRevA.52.2569} {\bibfield
  {journal} {\bibinfo  {journal} {Phys. Rev. A}\ }\textbf {\bibinfo {volume}
  {52}},\ \bibinfo {pages} {2569} (\bibinfo {year} {1995})}\BibitemShut
  {NoStop}%
\bibitem [{\citenamefont {Callen}(1985)}]{CallenHB1985book}%
  \BibitemOpen
  \bibfield  {author} {\bibinfo {author} {\bibfnamefont {H.~B.}\ \bibnamefont
  {Callen}},\ }\href@noop {} {\emph {\bibinfo {title} {Thermodynamics and an
  Introduction to Thermostatistics}}}\ (\bibinfo  {publisher} {Wiley, New
  York},\ \bibinfo {year} {1985})\BibitemShut {NoStop}%
\end{thebibliography}%

\end{document}